\documentclass[aps,prl,twocolumn,groupedaddress,amsmath]{revtex4}

\usepackage{graphics}
\usepackage{amssymb}

\newcommand*{\cD}{{\cal D}}

\newcommand*{\cG}{{\cal G}}

\newcommand*{\cT}{{\cal T}}

\begin{document}

\title{Noise and Counting Statistics of Insulating Phases in One-Dimensional Optical Lattices}

\author{Austen Lamacraft}
\affiliation{Rudolf Peierls Centre for Theoretical Physics, 1 Keble Road, Oxford OX1 3NP, UK
and All Souls College, Oxford.}
\date{\today}
\email{austen.lamacraft@all-souls.ox.ac.uk}

\begin{abstract}

We discuss the correlation properties of current carrying states of one-dimensional insulators, which could be realized by applying an impulse to atoms loaded onto an optical lattice. While the equilibrium noise has a gapped spectrum, the quantum uncertainty encoded in the amplitudes for the Zener process gives a zero frequency contribution out of equilibrium. We derive a general expression for the generating function of the full counting statistics and find that the particle transport obeys binomial statistics with doubled charge, resulting in super-Poissonian noise that originates from the bosonic nature of particle-hole pairs.

\end{abstract}


\maketitle

The characterization of particle fluxes using noise and full counting statistics (FCS) has played a very prominent role in quantum optics and, more recently, mesoscopic physics. Lately, these same ideas have come to the fore in the study of cold atomic gases, where noise measurements have been shown to be sensitive probes of inter-particle correlations~\cite{folling2005,greiner2005} as well as quantum statistics~\cite{ottl2005,schellekens2005}.

Although the idea of measuring the full distribution of particle transport has generated great theoretical interest recently for the case of massive particles (usually electrons)~\cite{levitov1993,nazarov2003}, the experimental determination in Ref.~\onlinecite{ottl2005} of the FCS for an atom laser represents the first such measurement. The observation of Poissonian statistics without bunching confirms that the atom laser shares many characteristics with its optical counterpart. 
%
%
This means that the noise spectrum
\begin{equation}
S(\omega)=\int dt\, e^{i\omega t}\langle\hat J(0)\hat J(t)+\hat J(t)\hat J(0)\rangle,
\end{equation}
is proportional to the current: $S(\omega)=2\langle\hat J\rangle$. The zero frequency contribution is usually called shot noise, and is characterized by a Fano factor $F\equiv S(0)/2\langle J\rangle$, here equal to 1.
\begin{figure} 
\begin{center}
\setlength{\unitlength}{3in}
\begin{picture}(1, 0.83)(0,0)
  \put(-0.05,0){\resizebox{1\unitlength}{!}{\includegraphics{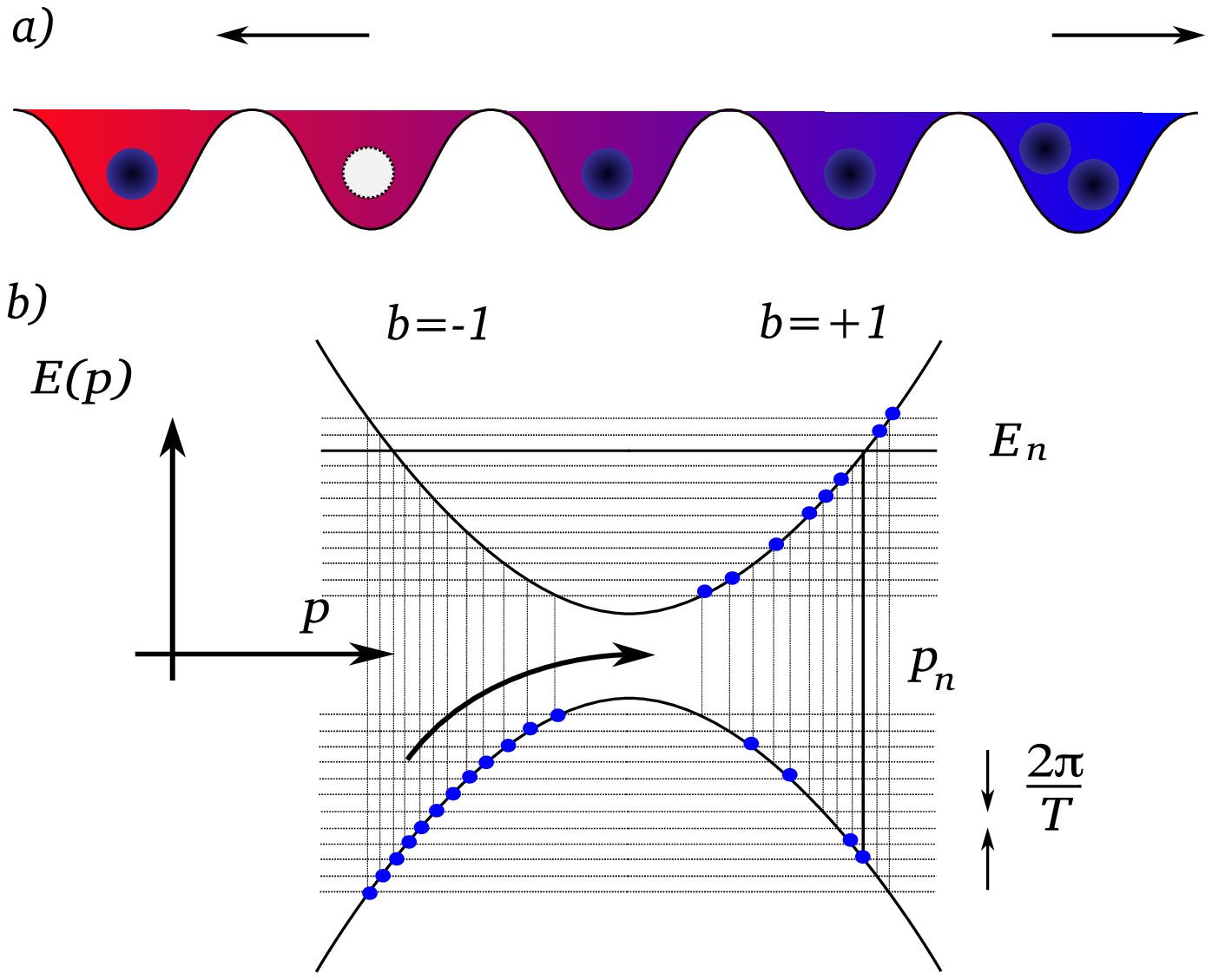}}}
  \end{picture}
\end{center}
\caption{a) Schematic view of particle-hole pair production in a one-dimensional insulator subject to a force. b) The discretization used to evaluate the counting statistics.
\label{fig:ph}}
\end{figure}
The FCS is characterized by the generating function $\chi(\lambda,T)$ for the moments of the number of particles transported in time $T$. Its logarithm generates the cumulants of the distribution. Thus for the Poisson distribution we have
\begin{equation*}
\ln\chi(\lambda,T)=\langle\hat J\rangle T\left(e^{i\lambda}-1\right).
\end{equation*}
Most discussions in mesoscopic physics concern the generation of shot noise from partitioning of the particle flux by localized scatterers. In this Letter we consider another natural source of shot noise: the Zener process by which current-carrying states are generated on a lattice~\cite{zener1934}. Zener tunneling has been observed in one-dimensional optical lattices loaded with cold atoms~\cite{morsch2001}, and this is a natural setting for the phenomena that we will discuss. Our aim is to extend Zener's result $-\left(\Phi/2\pi\hbar\right)\ln\left[1-P_{LZ}\right]$ for the transition rate per unit volume, where $P_{LZ}$ is the Landau-Zener probability and $\Phi$ the applied force, to the entire distribution of the resulting charge transport. 


When fermions are partitioned by a localized scatterer with transmission probability $\cT$, it was found that the Fano factor is $1-\cT$~\cite{lesovik1989}, i.e. the noise is suppressed relative to the Poisson case. In Ref.~\cite{levitov1993} this suppression was revealed to be a consequence of the binomial statistics of charge transport. In the present case, we might expect by analogy that when $P_{LZ}$ is small, Poissonian statistics is obtained with a Fano factor of one, falling to zero as the probability rises to unity. Such an expectation corresponds to `classical' picture of particle-hole creation, illustrated in Fig.~\ref{fig:ph}a). Remarkably, this expectation is only partially borne out. We find that the FCS resulting from an impulse applied to an insulating state on a one-dimensional lattice is binomial, but with doubled `charge'. This means that the Fano factor is $F=2\left(1-P_{LZ}\right)$. The origin of this unusual `bunching' behaviour, which can occur for bosons or fermions, is the coherent creation of particles and holes, or alternatively the bosonic behavior of particle-hole pairs.

There are two natural situations to consider. Fermions loaded onto a lattice form a band insulator at integer filling.
In the absence of interactions current-carrying states can be long-lived as there is no way for quasi-momentum to relax~\cite{ott2004}. 
Restricting ourselves to two bands, one empty and one full -- more bands present no difficulty -- we fix for concreteness the massive Dirac band structure near the Brillouin zone boundary 
\begin{equation}\label{Dirac}
H^B=c p\,\hat\sigma_3 +u\,\hat\sigma_1.
\end{equation}
Here $p$ is the deviation of the (quasi-)momentum from the Brillouin zone boundary, $c$ is a velocity, set to unity from now on, and $2u$ is the bandgap, proportional to the intensity of the laser light field that mixes left and right moving particles. In the basis of left and right movers, the two bands have the form
\begin{eqnarray}
\psi_+(p)&=&\frac{1}{\sqrt{2E_p(E_p-p)}}\left(\begin{array}{c}u \\E_p-p\end{array}\right)\nonumber\\
\psi_-(p)&=&\frac{1}{\sqrt{2E_p(E_p-p)}}\left(\begin{array}{c}p-E_p \\u\end{array}\right)
\end{eqnarray}
with energies $E_\pm(p)\equiv\pm E_p\equiv\pm\sqrt{p^2+u^2}$.  Introducing the corresponding creation and annihilation operators, the normal-ordered current operator at the origin can be expressed as
\begin{eqnarray} \label{band_current}
\hat J(0)&=&\sum_{p,p'}v^B_+(p,p')\hat\psi^\dagger_+(p)
\hat\psi^{\vphantom{\dagger}}_+(p')\nonumber\\
&&-v^B_-(p,p')
\hat\psi^{\vphantom{\dagger}}_-(p')\hat\psi^\dagger_-(p)\nonumber\\
&&+\left[w^B(p,p')\hat\psi^\dagger_+(p)
\hat\psi^{\vphantom{\dagger}}_-(p')+\mathrm{h.c.}\right]
\end{eqnarray}
(we work in units with $\hbar=1$) Here 
\begin{eqnarray} \label{band}
v^B_+(p,p')&=& \frac{u^2-\left(E_p-p\right)\left(E_{p'}-p'\right)}{2\sqrt{E_p(E_p-p)}\sqrt{E_{p'}(E_{p'}-p')}} \nonumber\\
v^B_-(p,p')&=&-v^B_+(p,p') \nonumber\\
w^B(p,p')&=& -\frac{u\left(E_p-p+E_{p'}-p'\right)}{2\sqrt{E_p(E_p-p)}\sqrt{E_{p'}(E_{p'}-p')}}
\end{eqnarray}
When $p=p'$ the band diagonal parts of the current Eq.~(\ref{band_current}) involve the group velocities $v^B_\pm(p,p)= dE_{\pm}(p)/dp$. 

We will also consider a one-dimensional bosonic Mott insulator. modeled by hardcore bosons $b_i^2=0$ on a lattice at half filling, with a staggered field
\begin{equation} \label{Mott}
H^M=\sum_i \left[b_i^\dagger b_{i+1}^{\vphantom{\dagger}}+\mathrm{h.c}\right]+u\left(-\right)^ib_i^\dagger b_i^{\vphantom{\dagger}}
\end{equation}
This model is equivalent under a Jordan-Wigner transformation to a band insulator of fermions with two bands. The expression for the current, found by standard methods~\cite{mahan1981}, has the same form as Eq.~(\ref{band_current}), with 
\begin{eqnarray} \label{Mott_current}
v^M_+(p,p')&=& \frac{\left(E_{p'}-u\right)\xi_p e^{ip'}+\left(E_p-u\right)\xi_{p'}e^{-ip}}{2\sqrt{E_p(E_p-u)}\sqrt{E_{p'}(E_{p'}-u)}}  \nonumber\\
v^M_-(p,p')&=&-v^M_+(p,p') \nonumber\\
w^M(p,p')&=& \frac{\xi_p\xi_{p'}e^{ip'}-\left(E_p-u\right)\left(E_{p'}-u\right)e^{-ip}}{2\sqrt{E_p(E_p-u)}\sqrt{E_{p'}(E_{p'}-u)}} ,
\end{eqnarray}
with $E_p=\sqrt{\xi_p+u^2}$ and $\xi_p=\sin p$, with $p$ the deviation from $\pi/2$, the zone boundary. For $p\ll 1$ Eq.~(\ref{Mott_current}) coincides with Eq.~(\ref{band}). The specific forms of $v_{\pm}(p,p')$ and $w(p,p')$ will not be needed. We now consider what happens when an impulse is applied to the lattice, that is, a time-dependent but spatially constant force $\Phi(t)$. The problem remains translationally invariant with each momentum state $p_i$ being mapped to one at $p_f=p_i+\int \Phi(t') dt'$. In general, there is some amplitude to make a transition to the upper band, meaning that the wavefunction of a given momentum state, in the basis of upper and lower bands, evolves as
\[\Psi_i(p_i)=\left(\begin{array}{c}0 \\ 1\end{array}\right)\to \Psi_f(p_f)=\left(\begin{array}{c}B_{p_f} \\ A_{p_f}\end{array}\right).\]
We now use these amplitudes to compute the resulting current fluctuations. Let us begin by finding the average current. Taking the expectation of the Heisenberg representation form of $\hat J(0,t)$
\begin{eqnarray}
\langle J(0,t)\rangle=\sum_p \left[v_+(p,p)-v_-(p,p)\right]|B_p|^2\nonumber\\+2w(p,p)\mathrm{Re}\,B_p^*A_p e^{i(E_+(p)-E_-(p)) t}.
\end{eqnarray}
The last term gives rise to a transient, and in the $t\to\infty$ limit we have the natural result
\begin{equation} \label{current}
\langle \hat J(0,t\to\infty)\rangle=\sum_p \left[\frac{dE_+(p)}{dp}-\frac{dE_-(p)}{dp}\right] |B_p|^2.
\end{equation}
Moving on to the current fluctuations, the expectation for the connected correlation function
\[\langle \hat J(0,t)\hat J(0,t')\rangle_c\equiv\langle \hat J(0,t)\hat J(0,t')\rangle-\langle \hat J(0,t)\rangle\langle\hat J(0,t')\rangle,\]
is straightforwardly obtained but rather lengthy. It is useful to first consider the equilibrium case when $A_p=1$, $B_p=0$. Then we have
\begin{equation} \label{quantum_noise}
\langle \hat J(0,t)\hat J(0,0)\rangle_c=\sum_{p,p'} w(p,p')w(p',p)e^{-i(E_+(p)-E_-(p'))t}.
\end{equation}
Note that this arises from off-diagonal part of the current Eq.~(\ref{band_current}). For the Dirac case the integrals are readily performed to give
\begin{eqnarray}
\langle \hat J(0,t)\hat J(0,0)\rangle_c&=&\frac{u^2}{2\pi^2}\left[K_1(-iu t)^2\right.\nonumber\\
&&\left.+\left(K_0(-iu t)+\pi H_0(u t)\right)^2\right],
\end{eqnarray}
where $K_n(x)$ is a Bessel function and $H_n(x)$ a Struve function. At $t\to 0$ we have the behavior
\[\langle \hat J(0,t)\hat J(0,0)\rangle_c\to -\frac{1}{2\pi^2(t-i0)^2}.\]
In this limit the correlator matches that of the massless system, which is fixed by the Schwinger commutators
\begin{eqnarray} 
\left[\hat J_{L/R}(x),\hat J_{L/R}(x')\right]&=&\pm\frac{i}{2\pi}\delta'(x-x')\nonumber\\
\left[\hat J_{L/R}(x),\hat J_{R/L}(x')\right]&=&0.
\end{eqnarray}
As may be readily seen from Eq.~(\ref{quantum_noise}), the noise spectrum contains only positive frequencies, indicating that the system can only absorb energy at zero temperature. Furthermore the spectrum is gapped at $2u$, meaning that there is no noise at zero frequency. 

It is thus reasonable to drop the gapped contributions to the noise (we will justify this more rigorously later). Additionally, we now focus on the particle-hole symmetric case where $E_{\pm}(p)=\pm E_p$, and $v_+(p,p')=-v_-(p,p')\equiv v(p,p')$, as in our two examples. Then the current correlator out of equilibrium is
\begin{widetext}
\begin{eqnarray} \label{non_eqm_noise}
\langle \hat J(0,t)\hat J(0,t')\rangle_c=
2\sum_{p,p'}& v(p,p') v(p',p) \left[e^{-i\left(E_p-E_{p'}\right)\left(t-t'\right)}|B^{\vphantom{*}}_p|^2|A^{\vphantom{*}}_{p'}|^2
+ e^{-i\left(E_p-E_{p'}\right)\left(t+t'\right)} B^*_{p}A^{\vphantom{*}}_{p}B^{\vphantom{*}}_{p'}A^*_{p'}\right]
\end{eqnarray}
\end{widetext}
Notice that the noise correlator is non-stationary, since the non-equilibrium state is not an eigenstate. It is still possible to define a Fano factor as $\langle Q_T^2\rangle_c=F\langle Q_T\rangle$, where $Q_T$ is the current passed in time $T$. Then Eq.~(\ref{current}) and Eq.~(\ref{non_eqm_noise}) give (we use the property $v(p,-p)=0$)
\begin{equation} \label{Fano}
F=\frac{2\sum_p |E'(p)||A_p|^2|B_p|^2}{\sum_p E'(p)|B_p|^2}.
\end{equation}
This result closely resembles the formula for the shot noise of a point contact. The key difference lies in the factor of two, so that in the limit $B_p\to 0$ we get $F=2$ rather than 1. The origin of this surprising super-Poissonian behavior is the second term of Eq.~(\ref{non_eqm_noise}).
   
We now discuss the FCS for this problem. Two toy models of `ideal' detectors have been considered in the literature~\cite{levitov1996,kindermann2003} -- both give the generating function of the FCS as
\begin{eqnarray} \label{FCS}
\chi(\lambda,T)=\Big\langle T_{c} \exp\left(\frac{i\lambda}{2}\int_{c(T)} dt' \mathrm{sgn}_{c}(t') \hat J(0,t')\right)\Big\rangle.
\end{eqnarray}
Here $T$ denotes the measurement time, and $c(T)$ is a (Keldysh) time contour that passes from $0$ to $T$ and back again. $T_c$ is the operation of time ordering on this contour, and $\mathrm{sgn}_c$ is $+1$ on the forward branch, and $-1$ on the backward branch. The expectation value is taken over the current-carrying state of the system. It should be noted that the measurement performed in Ref.~\onlinecite{ottl2005} would seem to have little in common with the idealized detection arrangements for which the above form of $\chi(\lambda,T)$ was found. We hope to provide a detailed analysis of this issue in a forthcoming publication: for the moment we will work with Eq.~(\ref{FCS}) for want of something better.
 
We use Wick's theorem to rewrite Eq.~(\ref{FCS}) in normal-ordered form
\begin{eqnarray}\label{normal}
\chi(\lambda,T)=\chi_{\mathrm{Vac}}(\lambda,T)\big\langle :\exp\left(-i\mathrm{Tr}\left[\cG\Lambda\hat\Psi^{\vphantom{\dagger}}\otimes\hat\Psi^{\dagger}\right]\right):\big\rangle\nonumber\\
\chi_{\mathrm{Vac}}(\lambda,T)=\exp\left[\mathrm{Tr}\ln\, \cG^{-1}\right],\qquad \cG=\left(\openone-i\Lambda\cD \right)^{-1}
\end{eqnarray}
In the above we have introduced the doublet
\[\hat\Psi^{\vphantom{\dagger}}_p=\left(\begin{array}{c} \hat\psi^{\vphantom{\dagger}}_+(p)\\\hat\psi^{\vphantom{\dagger}}_-(p)\end{array}\right),  \]
while $\cD_{p,p'}(t,t')\equiv T_c \hat\Psi^{\vphantom{\dagger}}_p(t)\otimes\hat\Psi^{\dagger}_{p'}(t')-:\hat\Psi^{\vphantom{\dagger}}_p(t)\otimes\hat\Psi^{\dagger}_{p'}(t'):$ is the Feynman propagator
\[ \cD_{p,p'}(t,t')=\delta_{p,p'}\left(\begin{array}{cc}\theta_c(t,t') & 0 \\0 & -\theta_c(t',t)\end{array}\right). \]
$\theta_c(t,t')$ is $1$ if $t$ comes after $t'$ on the contour, and zero otherwise.
%
%
%
The matrix $\Lambda_{p,p'}(t)$ is introduced so that
\[\frac{\lambda}{2}\mathrm{sgn}_c(t) \hat J(0,t)=\sum_{p,p'}\hat\Psi^{\dagger}_p\Lambda_{p.p'}(t)\hat\Psi^{\vphantom{\dagger}}_{p'}.\]
That is, it contains the $v_{\pm}(p,p')$ and $w(p,p')$ of Eq.~(\ref{band_current}). The `$\mathrm{Tr}$' in Eq.~(\ref{normal}) is a trace over the band, time, and $p$ indices.

In finding Eq.~(\ref{normal}) we have factored off the part of the generating function that exists in the `vacuum'. Recall that we earlier showed that the quantum noise Eq.~(\ref{quantum_noise}) in the ground state has a gapped spectrum. We thus do not expect it to contribute to the FCS at long times $T$. Since Eq.~(\ref{quantum_noise}) originated from the band off-diagonal contribution to the current, we can drop these effects at the outset by setting $w(p,p')=0$ in $\Lambda_{p,p'}$. 


Alternatively, we can perform a clean calculation in the following way~\cite{levitov1996}. We pass to a set of equally spaced states in energy space with energies $\pm E_n=\pm u\pm2\pi n/T$ (assuming also that $2u$ is an integer multiple of $2\pi/T$).  In this basis the current operator takes the form
\begin{widetext}
\begin{eqnarray} \label{energy_basis}
J(0,t)=\sum_{n,n',b,b'}\frac{v_{bb'}(n,n')}{|v_{bb}(n,n)v_{b'b'}(n',n')|^{1/2}}\left[\hat\psi^\dagger_{+bn}\hat\psi^{\vphantom{\dagger}}_{+b'n'}e^{-i\left(E_n-E_{n'}\right)t}+\hat\psi^{\vphantom{\dagger}}_{-b'n'}\hat\psi^\dagger_{-bn}e^{i\left(E_n-E_{n'}\right)t}\right],
\end{eqnarray}
The `branch' index $b=-1,1$ is needed to account for $p<0$ and $p>0$ states with momenta $b\,p_n$ such that $E_n=E(p_n)$ (see Fig.~\ref{fig:ph}b.). In Eq.~(\ref{energy_basis}) $v_{bb'}(n,n')=v(b\,p_n,b'\,p_{n'})$. 
The advantage of this approach is that not only do the effects of $w(p,p')$ disappear but $\chi(\lambda,T)$ takes a diagonal form with $\chi_{\mathrm{Vac}}\left(\lambda,T\right)=1$ and
%
\begin{eqnarray} \label{genFCS}
\chi(\lambda,T)=\Big\langle:\exp\Big(T\sum_{n,b}\hat\Psi^\dagger_{bn}\left(\begin{array}{cc}e^{i\lambda b}-1 & 0 \\0 &1-e^{i\lambda b}\end{array}\right)\hat\Psi_{bn}^{\vphantom{\dagger}}\Big):\Big\rangle,
\end{eqnarray}
%
%
Eq.~(\ref{genFCS}) represents a general form for $\chi(\lambda,T)$, ready for averaging over the state of the system. 
%
%
The state following an applied impulse is
\[\prod_{n,b} \left[A_{bn}+B_{bn}\hat\psi^{\dagger}_{+bn}\hat\psi^{\vphantom{\dagger}}_{-bn}\right]|0\rangle.\]
(note that the assumption of particle-hole symmetry enters here in an essential way -- pairing of particle and hole states at fixed momentum translates to pairing at given $n$) This state has
\begin{eqnarray} \label{normal_form}
\left(\begin{array}{cc}\langle \hat\psi^\dagger_{+bn}\hat\psi^{\vphantom{\dagger}}_{+bn}\rangle &\langle \hat\psi^\dagger_{+bn}\hat\psi^{\vphantom{\dagger}}_{-bn}\rangle \\\langle \hat\psi^\dagger_{-bn}\hat\psi^{\vphantom{\dagger}}_{+bn}\rangle &\langle \hat\psi^{\vphantom{\dagger}}_{-bn}\hat\psi^\dagger_{-bn}\rangle\end{array}\right)&=&\frac{1}{T}\left(\begin{array}{cc}|B^{\vphantom{*}}_{bn}|^2 & B^{\vphantom{*}}_{bn}A_{bn}^* \\A^{\vphantom{*}}_{bn}B_{bn}^* & -|B^{\vphantom{*}}_{bn}|^2\end{array}\right)
\end{eqnarray}
The average of the normal-ordered exponent in Eq.~(\ref{genFCS}) can then be found
\begin{eqnarray*}
\chi(\lambda,T)
&=&\prod_{n,b}\mathrm{det}\left(\begin{array}{cc}|A^{\vphantom{*}}_{bn}|^2+|B^{\vphantom{*}}_{bn}|^2 e^{i\lambda b}
 & B^{\vphantom{*}}_{bn}A^*_{bn}\left(e^{i\lambda b}-1\right) \\ A^{\vphantom{*}}_{bn}B^*_{bn}\left(1-e^{i\lambda b}\right) & |A^{\vphantom{*}}_{bn}|^2+|B^{\vphantom{*}}_{bn}|^2 e^{i\lambda b} \end{array}\right)\nonumber\\
 \ln\chi(\lambda,T)&=&\sum_{n,b}\ln\left[|A^{\vphantom{*}}_{bn}|^2+|B^{\vphantom{*}}_{bn}|^2e^{2i\lambda b}\right].
\end{eqnarray*}
\end{widetext}
We used the normalization condition $|A^{\vphantom{*}}_{bn}|^2+|B^{\vphantom{*}}_{bn}|^2=1$. To make things concrete, let's consider the case of a constant force $\Phi$ applied for time $\tau$, with the resulting impulse $\Phi\tau$ not creating particle-hole pairs too far away from the zone boundary, so that the Dirac structure Eq.~(\ref{Dirac}) is a good approximation. As long as $\Phi\tau\gg u/c$, however, most of the resulting particle-hole excitations are located on the linear part of the spectrum at $p>0$, so that we can neglect the crossing point and take the amplitudes to be given by the Landau-Zener formula
\begin{eqnarray*}
|B_{bn}|^2=\begin{cases}
0 & b=-1 \cr
P_{LZ}\equiv e^{-\frac{\pi u^2}{c \Phi}} & b=1,
\end{cases}
\end{eqnarray*}
giving the FCS
\begin{equation} \label{FCSfinal}
\chi(\lambda,T)=\left[\left(1-P_{LZ}\right)+P_{LZ}e^{2i\lambda}\right]^{ c\Phi\tau T/2\pi},
\end{equation}
and the Fano factor $F=2\left(1-P_{LZ}\right)$, in agreement with Eq.~(\ref{Fano}). The result Eq.~(\ref{FCSfinal}) corresponds to binomial statistics of particle transport, with a probability $P_{LZ}$ of transporting two charges (a particle passing in one direction and a hole in the other) per `attempt', and a number of attempts $c\Phi\tau T/2\pi$. Note that the unusual doubling is a consequence of the coherent creation of particles and holes: if the off-diagonal parts of Eq.~(\ref{normal_form}) are set to zero, the FCS is a product of binomial statistics for particles and holes separately, with Fano factor $\leq 1$. The classical picture of Fig~\ref{fig:ph}a) upon which our intuition was based corresponds to this case.

A brief comment is in order concerning the case without particle-hole symmetry. In general the second term in Eq.~(\ref{non_eqm_noise}) will give rise to a zero frequency contribution to the noise from a pair of momenta $p,p'$ with $E_+(p)+E_-(p')=0$. This contribution involves the combination of amplitudes $B^*_{p}A^{\vphantom{*}}_{p}B^{\vphantom{*}}_{p'}A^*_{p'}$, and is thus in general sensitive to the phase (the `Stokes phase'). One would expect a general decrease of the Fano factor for this reason, though super-Poissonian behavior is still possible. 

To conclude, we have investigated the counting statistics of the Zener process, and discovered an unusual super-Poissonian behavior. It would be interesting to perform such calculations for more realistic models of insulators, where particles and holes can interact, to see how the results are modified.

\bibliographystyle{apsrev}
\bibliography{FCS}
 
 \end{document}